\documentclass[nofootinbib,prd,showpacs]{revtex4-1}
\usepackage{amsmath}
\usepackage{amsfonts}
\usepackage{amssymb}
\usepackage{subfigure}
\usepackage{graphicx}
\graphicspath{ {Fig/} }
\usepackage[usenames,dvipsnames]{xcolor}
\usepackage{hyperref}   
\hypersetup{colorlinks=true, 
citecolor=MidnightBlue, linkcolor=MidnightBlue, urlcolor=Cyan}
\usepackage{tikz}
\definecolor{purple}{rgb}{1,0,1}
\definecolor{lime}{HTML}{A6CE39} % needs xcolor

\newcommand{\orcidicon}{%
	\begin{tikzpicture}
	\draw[lime, fill=lime] (0,0) 
		circle [radius=0.16] 
		node[white] {{\fontfamily{qag}\selectfont \tiny ID}};
	\draw[white, fill=white] (-0.0625,0.095) 
		circle [radius=0.007];
	\end{tikzpicture}	\hspace{-2mm}
}
\newcommand\orcidDaniele{{\href{https://orcid.org/0000-0003-4379-2549}{\orcidicon}}}
\newcommand\orcidFrancisco{{\href{https://orcid.org/0000-0002-9388-8373}{\orcidicon}}}
\begin{document}
%========================================================
\title{Effective  $f(R)$ actions for modified Loop Quantum Cosmologies via order reduction}
%=================================================================
\author{Ana Rita Ribeiro}
\email{anaritaribeiro95@hotmail.com}
\affiliation{Instituto de Astrof\'{i}sica e Ci\^{e}ncias do Espa\c{c}o, Faculdade de Ci\^encias da Universidade de Lisboa, Edif\'{i}cio C8, Campo Grande, P-1749-016, Lisbon, Portugal}
%=================================================================
\author{Daniele Vernieri\orcidDaniele\!\!}
\email{daniele.vernieri@unina.it}
\affiliation{Department of Physics ``E. Pancini'', University of Naples ``Federico II'', Naples, Italy,}
\affiliation{INFN Sez. di Napoli, Compl. Univ. di Monte S. Angelo, Edificio G, Via	Cinthia, I-80126, Naples, Italy.}

%=================================================================
\author{Francisco S. N. Lobo\orcidFrancisco\!\!}
\email{fslobo@fc.ul.pt}
\affiliation{Instituto de Astrof\'{i}sica e Ci\^{e}ncias do Espa\c{c}o, Faculdade de Ci\^encias da Universidade de Lisboa, Edif\'{i}cio C8, Campo Grande, P-1749-016, Lisbon, Portugal}
\affiliation{Departamento de F\'{i}sica, Faculdade de Ci\^{e}ncias, Universidade de Lisboa, Edifício C8, Campo Grande, PT1749-016 Lisbon, Portugal}

%=================================================================
%-----------------------------------------------------------------
\date{\LaTeX-ed \today}
%========================================================
\begin{abstract}
%========================================================
General Relativity is an extremely successful theory, at least for weak gravitational fields, however, it breaks down at very high energies, such as in correspondence of the initial singularity. Quantum Gravity is expected to provide more physical insights concerning this open question. Indeed, one alternative scenario to the Big Bang, that manages to completely avoid the singularity, is offered by Loop Quantum Cosmology (LQC), which predicts that the Universe undergoes a collapse to an expansion through a bounce. In this work, we use metric $f(R)$ gravity to reproduce the modified Friedmann equations which have been obtained in the context of modified loop quantum cosmologies. To achieve this, we apply an order reduction method to the $f(R)$ field equations, and obtain covariant effective actions that lead to a bounce, for specific models of modified LQC, considering matter as a scalar field. 
\end{abstract}
%========================================================
%\pacs{04.50.Kd,04.70.Bw}
%=================================================================
\maketitle
%=================================================================
\def\HMS{{\scriptscriptstyle{\rm HMS}}}
%========================================================
\bigskip
\hrule
\tableofcontents
\bigskip
\hrule
%========================================================
%\parindent0pt
%\parskip7pt
%\vspace{-10pt}
%========================================================

%%%%%%%%%%%%%%%%%%%%%%%%%%%%%%%%%%%%%%%%%%%%%%%%%%%%%%%%%%%%%%%%%
\section{Introduction}
%%%%%%%%%%%%%%%%%%%%%%%%%%%%%%%%%%%%%%%%%%%%%%%%%%%%%%%%%%%%%%%%%

%%%%%%%%%%%%%%%%%%%%%%%%%%%%%%%%%%%%%%%%%%%%%%%%%%%%%%%%%%%%%%%%%
\subsection{Motivations: Loop quantum cosmology and its modifications}
%%%%%%%%%%%%%%%%%%%%%%%%%%%%%%%%%%%%%%%%%%%%%%%%%%%%%%%%%%%%%%%%%

At the order of the Planck length, where General Relativity (GR) breaks down, a quantum theory of gravity is expected to provide insight on the behaviour of gravity at quantum scales~\cite{AmelinoCamelia:2008qg}, and consequently in addressing fundamental open questions~\cite{Carlip:2001wq}, such as the initial singularity problem. Loop Quantum Gravity (LQG)~\cite{Rovelli:1997yv,Thiemann:2007zz,Ashtekar:2004eh} is one of the main candidates, which predicts that classical differential geometry, at small space-time curvatures, is replaced by a discrete quantum geometry at the Planck scale. Indeed, Loop Quantum Cosmology (LQC), which is described as a symmetry-reduced model of LQG~\cite{Bojowald:2006da}, provides a more complete understanding of the initial singularity, and through an effective Hamiltonian description yields the following modified Friedmann equation~\cite{Bojowald:2001xe,Ashtekar:2006uz,Ashtekar:2006wn,Taveras:2008ke,Banerjee:2011qu,deHaro:2017yll,Singh:2006sg}:
\begin{equation}\label{LQC}
H^2 = \frac{1}{3}\kappa \rho \left(1-\frac{\rho}{\rho_c}\right)\,,
\end{equation}
where $H\equiv\dot{a}/a$ is the Hubble function, given in terms of the scale factor $a(t)$, and the critical energy density $\rho_c$ is defined as $\rho_c=\sqrt{3}/(32\pi^2 \gamma^3)$, with $\gamma \approx 0.2375$ the Barbero--Immirzi parameter~\cite{Meissner:2004ju}. Throughout this work, we use the geometrized system of natural units in which $c = \hbar = G = 1$ and, accordingly, $\kappa\equiv8\pi$.
The resulting equation of motion is just the usual Friedmann equation, for a spatially flat universe with a cosmological constant equal to zero, with a modified source, which has no extra degrees of freedom as compared to GR. The modified Friedmann equation~\eqref{LQC} dictates that the big bang singularity is replaced by a quantum bounce that occurs at the critical density $\rho_{c}$, determined by the underlying quantum geometry,  which is independent of the matter content of the universe, when $H=0$ and $\ddot{a} > 0$. 

However, the procedure used in LQC to obtain an effective Hamiltonian constraint does not grasp the totality of the full theory of LQG, since the Lorentzian term needs to be quantized in a different way than the Euclidean one, which leads to a different Hamiltonian constraint, for the spatially flat cosmological space-time. Thus, due to the quantization ambiguities in constructing the Hamiltonian constraint operator in isotropic loop quantum cosmology, this has the important implication that different treatments may arrive at different Planck scale physics.
One of the first attempts to understand this problem and to construct a Hamiltonian constraint that is actually similar to the construction in LQG, and inherits more features from the full theory, was undertaken in Ref.~\cite{Yang:2009fp}, in which Thiemann's regularization was used. 
Indeed, the Lorentz term of the gravitational Hamiltonian constraint in the spatially flat  Friedmann--Lema\^{i}tre--Robertson--Walker (FLRW) model was quantized by two approaches different from that of the Euclidean term. One of the approaches is very similar to the treatment of the Lorentz part of the Hamiltonian in loop quantum gravity and therefore inherits more features from the full theory. Two symmetric Hamiltonian constraint operators were constructed respectively in the improved scheme, and it was shown that both have the correct classical limit through the semiclassical analysis.
Nevertheless, there is still no systematic derivation of LQC as a cosmological sector of LQG and, as a  consequence, different paths to obtain LQC-like Hamiltonians from LQG followed~\cite{Dapor:2017rwv,Dapor:2017gdk,Alesci:2018qtm,Bilski:2019tji}. Due to these quantization ambiguities in constructing the effective Hamiltonian from LQG, it is pertinent to understand how different quantization methods affect the physical predictions. More specifically, there are still great possibilities for the expanding universe to recollapse due to quantum gravity effects.

Throughout this work, we will focus on two specific modified loop quantum cosmology (mLQC) models~\cite{Li:2018opr,Yang:2009fp}, which differ from standard LQC in the way the Lorentzian term in the Hamiltonian constraint is treated~\cite{Li:2019ipm}. One of the main goals of this study was to verify whether the big bang was still replaced by a quantum bounce. Moreover, numerical solutions were shown to be reliable approximations in order to extract phenomenological implications, particularly in the bounce regime.

%%%%%%%%%%%%%%%%%%%%%%%%%%%%%%%%%%%%%%%%%%%%%%%%%%%%%%%%%%%%%%%%%
\subsection{Modified loop quantum cosmology -- I: mLQC-I}
%%%%%%%%%%%%%%%%%%%%%%%%%%%%%%%%%%%%%%%%%%%%%%%%%%%%%%%%%%%%%%%%%

As emphasized in Ref.~\cite{Li:2018opr}, in LQC it is not necessary to eliminate the Lorentzian term explicitly in order to express the Hamiltonian constraint. In fact, if the Lorentzian term is maintained, different effective Hamiltonians are obtained. In this formal sense, the model derived in Ref.~\cite{Li:2018opr}, which here we denote by mLQC-I, by treating the Lorentzian term in the Hamiltonian constraint separately and using Thiemann's regularization~\cite{Yang:2009fp} yielded a more fundamental difference between LQC and mLQC-I, in that in the latter the evolution of the universe is described by two different branches, $b_+$ and $b_-$, and therefore it is asymmetric with respect to the bounce. 

Considering the $b_{-}$ branch, the modified Friedmann equation takes the form~\cite{Li:2018opr,Li:2019ipm}
\begin{equation}
H^{2}_- = \frac{\kappa\rho}{3}\left(1 - \frac{\rho}{\rho_{c}^{I}} \right)\left[ 1 + \frac{\gamma^{2} \rho / \rho_{c}^{I}}{(\gamma^{2}+1)\left(1+\sqrt{1-\rho/\rho_{c}^{I}}\right)^{2}}  \right] \,,
\label{H2I}
\end{equation}
where $\rho_{c}^{I} \equiv \rho_{c}/[4(1+\gamma^{2})]$ is the critical density at which the quantum bounce occurs, that is suppressed by a factor of $4(1+\gamma^{2})$ with respect to the standard model of LQC.
For $\rho/\rho_{c}^{I}\ll 1$, the standard Friedmann equation $H^{2} \approx \kappa \rho/3$ is recovered. 

Concerning the $b_{+}$ branch, the modified Friedmann equation takes the form~\cite{Li:2018opr,Li:2019ipm}
\begin{equation}
H^{2} _+ = \frac{8\pi G \alpha \rho_{\Lambda}}{3}\left(1 - \frac{\rho}{\rho_{c}^{I}} \right)\left[ 1 + \frac{\rho\left(1-2\gamma^{2}+\sqrt{1-\rho/\rho_{c}^{I}}\right)}{4\gamma^{2}\rho_{c}^{I}\left(1+\sqrt{1-\rho/\rho_{c}^{I}}\right)}  \right] \,,
\label{H2I2}
\end{equation}
where $\alpha \equiv (1 - 5\gamma^{2})/(\gamma^{2} + 1)$, $\rho_{\Lambda} \equiv 3/[8\pi G\alpha\lambda^{2}(1+\gamma^{2})^{2}]$, and $\lambda^{2} \equiv 4\sqrt{3}\pi \gamma$. For this branch, the limit of Eq.~\eqref{H2I2} when $\rho/\rho_{c}\ll 1$, is given by $H^{2} \approx 8\pi G_{\alpha}(\rho + \rho_{\Lambda})/3 $, where $G_{\alpha}:=\alpha G$. It can be shown that the classical energy conservation law still holds in this modified model of LQC, concerning both branches.

Taking the current cosmological constraints into account, the evolution of the universe must start from the $b_{+}$ branch, in the pre-bounce phase, and afterwards the evolution will be switched to the one described by the $b_{-}$ branch. In LQC, we only have one branch due to the fact that the Lorentzian term is not explicit in the Hamiltonian. The two branches coincide at the bounce, $\rho = \rho_{c}^{I}$~\cite{Li:2018opr}.
Contrary to the previous branch, in the $b_{+}$ branch, the quantum geometric effects are still present in the above limit. Comparing with the standard Friedmann equation, we see that these effects are present in the form of a modified Newtonian constant, $G_{\alpha}$, and an emergent positive cosmological constant, $\rho_{\Lambda}$. These pose phenomenological problems, if this branch lies in the post-bounce universe in which we live~\cite{Li:2018opr}.

%%%%%%%%%%%%%%%%%%%%%%%%%%%%%%%%%%%%%%%%%%%%%%%%%%%%%%%%%%%%%%%%%
\subsection{Modified loop quantum cosmology -- II: mLQC-II}
%%%%%%%%%%%%%%%%%%%%%%%%%%%%%%%%%%%%%%%%%%%%%%%%%%%%%%%%%%%%%%%%%

The second modification of LQC that we consider in this work, outlined in Ref.~\cite{Yang:2009fp}, was derived by using Thiemann's regularization, similar to the previous model, while using the proportionality of Ashtekar's connection and extrinsic curvature~\cite{Li:2018fco}. 
The corresponding modified Friedmann equation takes the form~\cite{Yang:2009fp}
\begin{equation}
H^2=\frac{2\kappa\rho}{3}\left(1-\frac{\rho}{\rho_{c}^{II}}\right)\left[\frac{1+4\gamma^2\left(\gamma^2+1\right)\rho/\rho_{c}^{II}}{1+2\gamma^2\rho/\rho_{c}^{II}+\sqrt{1+4\gamma^2\left(1+\gamma^2\right)\rho/\rho_{c}^{II}}}\right] \,,
\label{H2II}
\end{equation}
where we find that the quantum bounce occurs at the critical density $\rho_{c}^{II} \equiv 4(\gamma^{2} + 1)\rho_{c}$. Unlike the mLQC-I model, the bounce in this model is perfectly symmetric, as in LQC. 
However, the modified dynamics in the Planck regime are less trivial, similarly to the mLQC-I model. Another similarity between LQC and mLQC-II, is their classical limit. The limit of Eq.~\eqref{H2II}, for $\rho/\rho_{c}^{II}\ll 1$, is given by $H^{2} \approx \kappa\rho/3$~\cite{Li:2018fco}. Once again, using the modified equations that this model provides, the energy-momentum conservation law still holds without any change in the properties of the equation of state.

%%%%%%%%%%%%%%%%%%%%%%%%%%%%%%%%%%%%%%%%%%%%%%%%%%%%%%%%%%%%%%%%%
\subsection{Order reduction}
%%%%%%%%%%%%%%%%%%%%%%%%%%%%%%%%%%%%%%%%%%%%%%%%%%%%%%%%%%%%%%%%%

Now, the only covariant action that leads to second-order equations, under metric variation, is the Einstein--Hilbert (EH) action, and any change in this action leads to higher-order equations. Considering an action other than the one from GR, thus appears to be unjustified, given that it implies a departure of a framework that has proven to be successful in describing several physical phenomena. However, considering such theories might be useful, especially if they are close to the theory which we believe to be presently more correct. In fact, given that the modified Friedmann equations~\eqref{H2I}--\eqref{H2II} are obtained from modified sources, a reasonable question to ask is whether it is possible to derive them from an effective action, other than the EH action. 
An interesting analysis was considered in the Palatini approach of $f(R)$ gravity in Ref.~\cite{Olmo:2008nf}. In this context, the functional Palatini $f(R)$ is a function of the trace of the energy-momentum tensor $T$, so that the modified Friedmann equation in Palatini $f(R)$ gravity does not involve any higher derivatives of the geometrical quantities and is just a function of the matter sources~\cite{Olmo:2011uz}. Indeed, this result provided new insights on the continuum properties of the discrete structure of quantum geometry. In this work, in order to deduce effective actions we use metric $f(R)$ gravity~\cite{Sotiriou:2008rp,Lobo:2008sg,DeFelice:2010aj,Nojiri:2010wj,Clifton:2011jh,Capozziello:2011et,Harko:2018ayt}, where the full field equations in this framework are of fourth-order. In this sense, we can either consider this theory as an exact theory, meaning that their field equations are considered as genuinely higher-order, or we can treat it as an effective field theory by only considering the solutions which are perturbatively close to GR as the physical ones and the rest as spurious~\cite{Bel:1985zz,Simon:1990ic,Simon:1991bm}.  A method that provides the latter approach is that of covariant order reduction, which allows one to find solutions that are perturbatively close to GR.
A further advantage of this procedure consists in the fact that it may also serve as a model selection approach.

Indeed, it was shown that, at least for isotropic models in LQC, using a covariant order reduction then a covariant effective action can be found if one considers higher-order theories of gravity but faithfully follows effective field theory techniques~\cite{Sotiriou:2008ya}.
In this context, these techniques were applied within bouncing cosmolologies, which can be envisaged as candidates for solving the big bang initial singularity problem. 
In Ref.~\cite{Terrucha:2019jpm}, bouncing solutions in a modified Gauss--Bonnet gravity theory were explored, of the type $R+f(G)$, where $R$ is the Ricci scalar, $G$ is the Gauss--Bonnet term, and $f$ some function of it~\cite{Terrucha:2019jpm}. In finding such a bouncing solution, the order reduction technique was used to reduce the order of the differential equations of the $R+f(G)$ theory to second order equations. As GR is a theory whose equations are of second order, this order reduction technique enables one to find solutions which are perturbatively close to GR. Furthermore, the covariant action of the order reduced theory was also obtained.
The analysis explored in Ref.~\cite{Terrucha:2019jpm} was extended in the context of $f(R,G)$ gravity by using the same order reduction technique~\cite{Barros:2019pvc}. Indeed, several covariant gravitational actions leading to a bounce are directly selected by demanding that the Friedmann equation derived within such gravity theories coincides with the one emerging from LQC. The same approach has also been implemented in the context of $f(Q)$ symmetric teleparallel gravity theories, where $Q$ is the non-metricity scalar~\cite{Bajardi:2020fxh}.

%%%%%%%%%%%%%%%%%%%%%%%%%%%%%%%%%%%%%%%%%%%%%%%%%%%%%%%%%%%%%%%%%
\subsection{Outline of the paper}
%%%%%%%%%%%%%%%%%%%%%%%%%%%%%%%%%%%%%%%%%%%%%%%%%%%%%%%%%%%%%%%%%

In this work, we are interested in extending the analysis outlined in Ref.~\cite{Sotiriou:2008ya}, by applying the method of covariant order reduction to the modified Friedmann equations~\eqref{H2I}--\eqref{H2II}, within the modified loop quantum cosmologies considered above. To this effect, we write out the Lagrangian density of $f(R)$ gravity as the sum of the gravitational Lagrangian of GR and a deviation term, and present the final form of the reduced modified Friedmann equations.
In Sec.~\ref{SecII}, we present the covariant order reduction technique in some detail and outline a strategy to deduce the effective actions. In Sec.~\ref{SecIII}, we obtain the effective actions for the modified loop quantum cosmological models considered above. Finally, in Sec.~\ref{Sec:Conclusions}, we conclude and discuss our results.

%%%%%%%%%%%%%%%%%%%%%%%%%%%%%%%%%%%%%%%%%%%%%%%%%%%%%%%%%%%%%%%%%
\section{Covariant order reduction technique}\label{SecII}
%%%%%%%%%%%%%%%%%%%%%%%%%%%%%%%%%%%%%%%%%%%%%%%%%%%%%%%%%%%%%%%%%

%%%%%%%%%%%%%%%%%%%%%%%%%%%%%%%%%%%%%%%%%%%%%%%%%%%%%%%%%%%%%%%%%
\subsection{Reduced equations in $f(R)$ gravity}
%%%%%%%%%%%%%%%%%%%%%%%%%%%%%%%%%%%%%%%%%%%%%%%%%%%%%%%%%%%%%%%%%

In this work, we consider metric $f(R)$ gravity, whose corresponding action is given by
\begin{equation}
S = \frac{1}{2\kappa}\int_\mathcal{V} d^4x \sqrt{-g} f(R) + S_M (g_{\mu\nu},\Psi)\,,
\label{actionfR}
\end{equation}
where $g$ is the determinant of the metric tensor $g_{\mu\nu}$, $S_M$ is the matter action defined as $S_M=\int_\mathcal{V} d^4x \sqrt{-g}\, {\cal{L}}_M(g_{\mu\nu},\Psi)$, being ${\cal{L}}_M$ the matter Lagrangian density, in which matter is minimally coupled to the metric $g_{\mu\nu}$, and $\Psi$ collectively denotes the matter fields.
Without loss of generality, we parametrize the Lagrangian density $f(R)$ as the sum of the gravitational Lagrangian of GR and a deviation term, as follows
\begin{equation}
f(R)=R+2\Lambda+\epsilon \varphi (R)\,,
\label{fRparametrization}
\end{equation}
where $\Lambda$ is the cosmological constant, $\epsilon$ is a dimensionless parameter, and $\epsilon \varphi (R)$ represents the deviation from GR. Substituting Eq.~\eqref{fRparametrization} in Eq.~\eqref{actionfR}, and varying the action with respect to the metric yields the following gravitational field equations
\begin{equation}
G_{\mu\nu}-g_{\mu\nu}\Lambda+\epsilon\left[-\frac{1}{2}g_{\mu\nu}\varphi(R)+ \varphi^{\prime} (R)R_{\mu\nu}- \left(\nabla_\mu \nabla_\nu-g_{\mu\nu}\Box \right)\varphi^{\prime}(R)\right]=\kappa T_{\mu\nu}\,.
\label{ModFriedmann}
\end{equation}
The energy-momentum tensor $T_{\mu\nu}$ is defined as 
\begin{equation}
T_{\mu\nu}= - \frac{2}{\sqrt{-g}}\,\frac{\delta (\sqrt{-g} {\cal{L}}_M)}{\delta g^{\mu\nu}}\,.
\end{equation} 

We now apply the order reduction technique to Eq.~\eqref{ModFriedmann}. In the present case, this amounts to replacing the Ricci scalar and the Ricci tensor, in the terms of order $\epsilon$, by the expression we get for them, from the $\epsilon=0$ version of the same equations. Thus, setting $\epsilon=0$ in Eq.~\eqref{ModFriedmann}, yields the following
\begin{align}
R_{\mu\nu}^T = \frac{1}{2}g_{\mu\nu}(R_T+2\Lambda) + \kappa T_{\mu\nu} \label{eq2}
\end{align}
as the reduced expression of the Ricci tensor, where $R_T$ is the order reduced Ricci scalar. Taking into account the trace of Eq.~\eqref{eq2}, the latter provides $R_{T}=-\kappa T - 4 \Lambda$, and after substituting it into Eq.~\eqref{eq2}, we arrive at
\begin{align}
 R^{T}_{\mu\nu}=\kappa T_{\mu\nu}-\frac{1}{2}g_{\mu\nu}\kappa T -\Lambda g_{\mu\nu}\label{eq3}\,.
\end{align}
Replacing $R$ and $R_{\mu\nu}$, in the $\epsilon$-order terms in Eq.~\eqref{ModFriedmann}, with the above expressions, we finally get 
\begin{equation}
G_{\mu\nu} - g_{\mu\nu}\Lambda + \epsilon\left[-\frac{1}{2}g_{\mu\nu}\varphi(R_T) + \varphi^{\prime} (R_T)\left(\kappa T_{\mu\nu}-\frac{1}{2}g_{\mu\nu}\kappa T - \Lambda g_{\mu\nu}\right)
-\left(\nabla_\mu \nabla_\nu - g_{\mu\nu}\Box \right)\varphi^{\prime}(R_T)\right]=\kappa T_{\mu\nu}\,.
\label{ModFriedmann2}
\end{equation}
These are reduced field equations in the sense that we are not using the complete field equations to determine expressions for $R$ and $R_{\mu\nu}$. We should notice that this approximation is only valid when 
\begin{equation}
|\epsilon \varphi(R)| \ll |R|\,,
\label{condition}
\end{equation}
which has to be verified {\it a-posteriori} after an expression for $\varphi(R)$ is determined.

%%%%%%%%%%%%%%%%%%%%%%%%%%%%%%%%%%%%%%%%%%%%%%%%%%%%%%%%%%%%%%%%%
%\subsection{Time-time component of the reduced equations}
%%%%%%%%%%%%%%%%%%%%%%%%%%%%%%%%%%%%%%%%%%%%%%%%%%%%%%%%%%%%%%%%%

Throughout this work, we consider an isotropic and homogeneous universe described by the FLRW metric with a $(-,+,+,+)$ signature, which in spherical coordinates is given by 
\begin{equation}
ds^2=-dt^2+a(t)^2\bigg[\frac{dr^2}{1-kr^2}+r^2d\theta^2+r^2\sin^2\theta d\phi^2\bigg]\,,
\label{FLRW}
\end{equation}
where $a(t)$ is the scale factor and $k$ is the curvature of the universe, which can be set equal to $-1$, $0$ or $+1$, depending if one is considering a hyperspherical, spatially flat or hyperbolic universe, respectively. We assume a perfect fluid description for the content of the universe, given by the energy-momentum tensor, $T_{\mu\nu}=(\rho+p)U_{\mu}U_{\nu}+ p g_{\mu\nu}$, where $U^\mu$ is the four-velocity field of an observer comoving with the fluid, defined in such a way that $U_\mu U^\mu=-1$, and $\rho=\rho(t)$ and $p=p(t)$ are the fluid's energy density and isotropic pressure, respectively.

Now, taking into account the FLRW metric (\ref{FLRW}), then Eq.~\eqref{ModFriedmann2} provides the following modified Friedmann equation
\begin{equation}\label{eq8}
H^2=\frac{1}{3}\kappa \rho-\frac{k}{a^2}-\frac{\Lambda}{3}-\frac{\epsilon}{3}\left[\frac{1}{2}(3w+1)\varphi_{T}^{\prime}\kappa \rho +\varphi_{T}^{\prime}\Lambda +\frac{1}{2} \varphi_{T} + 9H^2 \varphi_{T}^{\prime\prime}  (1+w)(3w -1)\kappa\rho\right]\,,
\end{equation}
by taking into account a barotropic equation of state (EoS) $p=w\rho$, with $-1 \leq w \leq 1$.
Note that we recover the standard Friedmann equation for $\epsilon=0$. 

In the context of the order reduction method, we substitute the $H^{2}$ term within the $\epsilon$ term on the right hand side of Eq.~\eqref{eq8} by the standard Friedmann equation, which yields~\cite{Sotiriou:2008ya}
\begin{eqnarray}
H^2&=&\frac{1}{3}\kappa \rho-\frac{k}{a^2}-\frac{\Lambda}{3}-\frac{\epsilon}{3}\bigg[\frac{1}{2}(3w+1)\varphi^{\prime}(R_T)\kappa \rho +\varphi^{\prime}(R_T)\Lambda +\frac{1}{2} \varphi (R_T)
\nonumber\\
&&-\left(3\kappa \rho -\frac{9k}{a^2}-3\Lambda\right) \varphi^{\prime\prime}(R_T) (1+w)(1-3w)\kappa\rho\bigg]\,,
\label{mFE}
\end{eqnarray}
as our final form of the reduced modified Friedmann equation in $f(R)$ gravity. It is clear now, that the reduced differential equation is effectively a second-order one. With this result, we aim to find a function $\varphi(R_T)$, such that Eq.~\eqref{mFE} is the same as some Friedmann equation with a modified source.

%%%%%%%%%%%%%%%%%%%%%%%%%%%%%%%%%%%%%%%%%%%%%%%%%%%%%%%%%%%%%%%%%
\subsection{Strategy to obtain effective actions}
%%%%%%%%%%%%%%%%%%%%%%%%%%%%%%%%%%%%%%%%%%%%%%%%%%%%%%%%%%%%%%%%%

Here, we specify the conditions in which we will be using Eq.~\eqref{mFE}. In this context, we consider a spatially flat universe ($k = 0$) with a zero cosmological constant ($\Lambda = 0$), so that $R_{T}=-\kappa \rho(3w -1)$. Taking these aspects into account, then Eq.~\eqref{mFE} becomes
\begin{equation}
H^2=\frac{1}{3}\kappa \rho - \frac{\epsilon}{3}\left[\frac{1}{2}(3w+1)\varphi^{\prime}(R_{T})\kappa \rho  + \frac{1}{2} \varphi (R_{T}) - 3\kappa^{2} \rho^{2} \varphi^{\prime\prime}(R_{T}) (1+w)(1-3w)\right] \,.
\label{mFE2}
\end{equation}

In the following sections, we will compare Eq.~\eqref{mFE2} to the modified Friedmann equations~\eqref{H2I}--\eqref{H2II} considered in the modified loop quantum cosmologies, outlined in the Introduction. However, we reinforce the idea that there are two kinds of modified Friedmann equations at play. One of them is Eq.~\eqref{mFE2}, which was derived in the context of metric $f(R)$ gravity, and the other one was derived in the context of LQC. Thus, our main goal is to determine $\varphi(R_T)$ such that Eq.~\eqref{mFE2} is the same as Eqs.~\eqref{H2I}--\eqref{H2II}. In doing so, we will determine an effective action that leads to the quantum bounce which characterizes those models. Thus, the modified Friedmann equation, in the context of LQC and its modifications, can be written in a general way as
\begin{equation}
H^{2} = \frac{1}{3}\kappa \rho + \Psi (\rho) \,,
\label{DefPsirho}
\end{equation}
where $\Psi(\rho)$ is some algebraic function, which will depend on the model. Comparing with Eq.~\eqref{mFE2}, the requirement is that $\varphi (R_T)$ satisfies
\begin{equation}
-\frac{\epsilon}{3} \left[ \frac{1}{2}(3w+1)\varphi^{\prime}(R_{T})\kappa \rho  + \frac{1}{2} \varphi (R_{T}) - 3\kappa^{2} \rho^{2} \varphi^{\prime\prime}(R_{T}) (1+w)(1-3w) \right] = \Psi(\rho) \,,
\label{solve1}
\end{equation}
for a given $\Psi(\rho)$. The same equation, written in terms of $R_{T}$, using $R_{T}=-\kappa \rho(3w -1)$, is given by
\begin{equation}
-\frac{\epsilon}{3} \left[\frac{1}{2} \varphi (R_{T}) - \frac{(3w+1)}{2(3w-1)} R_{T}\, \varphi^{\prime}(R_{T}) + \frac{3(1+w)}{(3w-1)} R_{T}^{2} \, \varphi^{\prime\prime}(R_{T}) \right] = \Psi(R_{T})\,.
\label{GeneralwEqu}
\end{equation}

For each model, mLQC-I and mLQC-II, we will be considering the scenario of matter as a scalar field ($w=1$). An effective action for LQC was already determined in Ref.~\cite{Sotiriou:2008ya}, in the context of matter as a scalar field.

%%%%%%%%%%%%%%%%%%%%%%%%%%%%%%%%%%%%%%%%%%%%%%%%%%%%%%%%%%%%%%%%%
\section{Effective actions for modified loop quantum cosmology models}\label{SecIII}
%%%%%%%%%%%%%%%%%%%%%%%%%%%%%%%%%%%%%%%%%%%%%%%%%%%%%%%%%%%%%%%%%

In order to be in agreement with the approach leading to Eq.~\eqref{LQC}, we consider matter as a scalar field, i.e., we set the EoS parameter to be $w=1$. Therefore, Eq.~\eqref{mFE2} simplifies to
\begin{equation}
H^2=\frac{1}{3}\kappa \rho - \frac{\epsilon}{3}\left[2\varphi^{\prime}(R_{T})\kappa \rho  + \frac{1}{2} \varphi (R_{T}) + 12\kappa^{2} \rho^{2} \varphi^{\prime\prime}(R_{T})\right] \,,
\label{mFE3}
\end{equation}
and, as a consequence, Eq.~\eqref{solve1} reduces to
\begin{equation}
- \frac{\epsilon}{3} \left[ 2 \varphi^{\prime}(R_{T})\kappa \rho + \frac{1}{2} \varphi (R_{T}) + 12\kappa^{2} \rho^{2} \varphi^{\prime\prime}(R_{T}) \right] = \Psi(\rho) \,,
\label{solve2}
\end{equation}
with 
\begin{equation}
R_{T}=-2\kappa\rho \,.
\label{RmatterScalar}
\end{equation}
Note that $R_{T}$ is negative, as the energy density $\rho$ is positive. Finally, we write out Eq.~\eqref{solve2} in terms of $R_{T}$ as
\begin{equation}
- \frac{\epsilon}{3}\left[ \frac{1}{2} \varphi (R_{T}) - R_{T} \varphi^{\prime}(R_{T}) + 3 R_{T}^{2}  \varphi^{\prime\prime}(R_{T}) \right] = \Psi(R_{T})\,.
\label{solve3}
\end{equation}

Independently of the model of LQC that we are considering, at a given moment, the solution to the non-homogeneous equation~\eqref{solve3} consists of the sum of the solution to the corresponding homogeneous equation and its particular solution,
\begin{equation}
\varphi (R_{T}) = \varphi_{h} (R_{T}) + \varphi_{p} (R_{T})\,.
\label{handp}
\end{equation}
As such, before specifying any of the models, by substituting $\Psi(R_{T})$ in Eq.~\eqref{solve3} for each case, let us consider the homogeneous equation first, which is common to all cases and is given by
\begin{equation}
\frac{1}{2} \varphi (R_{T}) - R_{T} \varphi^{\prime}(R_{T}) + 3 R_{T}^{2}  \varphi^{\prime\prime}(R_{T}) = 0 \,.
\label{solve4}
\end{equation}
This equation is a second order Cauchy--Euler equation, having the form
\begin{equation}
x^2 \frac{d^2y(x)}{dx^2}+a x \frac{dy(x)}{dx} +by(x)=0\,.
\label{cauchy}
\end{equation}
Assuming a trial solution $y(x)=x^m$, we arrive at a characteristic equation for $m$, given by
$m^2 +(a-1)m+b=0$.
Through the roots of this equation, we obtain a solution for Eq.~\eqref{cauchy} whose form depends on the number of roots, and if they are real or complex. In the present case, the characteristic equation for the homogeneous Eq.~\eqref{solve4} is $m^2-4m/3+1/6=0$
and, solving for $m$, one gets two real roots, $m=\frac{1}{6}(4\pm \sqrt{10})$, meaning that the solution to Eq.~\eqref{solve4} is given by
\begin{equation}\label{Homo}
\varphi_{h} (R)=c_1 R^{\frac{1}{6}(4-\sqrt{10})}+c_2 R^{\frac{1}{6} (4+\sqrt{10})}\,.
\end{equation}
This solution does not contain analytic functions of $R$, meaning it is not locally given by a convergent power series. For this reason, and also for the fact that $\varphi_{h}(R)$ does not contribute to the full Eq.~\eqref{solve3}, we can set $c_1=c_2=0$, without loss of generality~\cite{Sotiriou:2008ya}. As such, the solutions to Eq.~\eqref{solve3} are simply given by the particular solution in question
\begin{equation}
\varphi (R_{T}) = \varphi_{p} (R_{T})\,,
\end{equation}
depending on $\Psi(R_{T})$.

%%%%%%%%%%%%%%%%%%%%%%%%%%%%%%%%%%%%%%%%%%%%%%%%%%%%%%%%%%%%%%%%%
\subsection{Effective action for LQC}
%%%%%%%%%%%%%%%%%%%%%%%%%%%%%%%%%%%%%%%%%%%%%%%%%%%%%%%%%%%%%%%%%

Here, for self-completeness and self-consistency, we briefly present the particular scenario studied in Ref.~\cite{Sotiriou:2008ya}, where an effective action for LQC was found, requiring matter to be a scalar field, in order to be in agreement with the approach leading to Eq.~\eqref{LQC}~\cite{Singh:2006sg}. From Eq.~\eqref{DefPsirho}, for LQC we have that
$\Psi(\rho) = - \kappa \rho^{2}/(3\rho_{c})$ or, as a function of $R_{T}$, using $R_{T}=-2\kappa\rho$, we have $\Psi(R_{T}) = - R_{T}^{2}/(12\kappa\rho_{c})$.
Therefore, in this case, Eq.~\eqref{solve3} is given by
\begin{equation}
\frac{1}{2} \varphi (R_{T}) - R_{T}\varphi^{\prime}(R_{T}) + 3R_{T}^{2}  \varphi^{\prime\prime}(R_{T}) = \frac{R_{T}^2}{4\kappa\rho_{c} \epsilon}\,.
\label{solveR1}
\end{equation}
Since we already covered the solution to Eq.~\eqref{solve4}, we now focus on its particular solution. Looking at the right-hand side of Eq.~\eqref{solveR1}, the particular solution should be of the form $\varphi_{p} (R_T) = A R_{T}^2$, where $A$ is a constant. Plugging this in Eq.~\eqref{solveR1}, one gets~\cite{Sotiriou:2008ya}:
\begin{equation}
\epsilon\varphi_{p} (R)=\frac{R^2}{18\kappa\rho_c}\,,
\label{deviationLQC}
\end{equation}
where we have dropped the subscript $T$ since, to $\epsilon$ order, $f(R)$ and $f(R_T)$ are the same. Thus, the $f(R)$ function for the effective action, that leads to Eq.~\eqref{LQC}, is~\cite{Sotiriou:2008ya}
\begin{equation}
f(R)=R+\epsilon\varphi_{p}(R)=R+\frac{R^2}{18\kappa\rho_c}+...\,,
\label{fRLQC}
\end{equation}
where the dots are an indication of an infinite series. 
Notice that the case of a quadratic model $f(R) \propto R^2$, in the context of LQC, was also considered in Ref.~\cite{Amoros:2014tha}.
An effective brane-world and the LQC background expansion histories were also reproduced from a modified gravity perspective in terms of a quadratic term in the Ricci scalar, in a theory non-minimally coupled with the matter Lagrangian~\cite{Olmo:2014sra}.
We can neglect the rest of the terms, as they are subdominant with respect to the quadratic term. According to the order reduction method, this solution is only valid for a certain range of curvatures, given by Eq.~\eqref{condition}, or equivalently, for a certain range of energy density values. In this case, the solution is valid for
\begin{equation}
R \gg -18\kappa \rho_c \quad \Rightarrow \quad \rho \ll 9\rho_{c}\,.
\label{LQCinterval}
\end{equation}
Thus, a metric $f(R)$ action was found in Ref.~\cite{Sotiriou:2008ya} which, when treated as an effective action, leads to the modified Friedmann equation, provided by LQC. Thus, we follow an analogous procedure for the modified models of LQC.

%%%%%%%%%%%%%%%%%%%%%%%%%%%%%%%%%%%%%%%%%%%%%%%%%%%%%%%%%%%%%%%%%
\subsection{Effective actions for mLQC-I}
%%%%%%%%%%%%%%%%%%%%%%%%%%%%%%%%%%%%%%%%%%%%%%%%%%%%%%%%%%%%%%%%%

As described in the Introduction, mLQC-I is divided in two branches. As such, we consider each of them separately and find an effective action for both.

%%%%%%%%%%%%%%%%%%%%%%%%%%%%%%%%%%%%%%%%%%%%%%%%%%%%%%%%%%%%%%%%%
\subsubsection{Effective action for the $b_{-}$ branch}
%%%%%%%%%%%%%%%%%%%%%%%%%%%%%%%%%%%%%%%%%%%%%%%%%%%%%%%%%%%%%%%%%

Considering Eq.~\eqref{H2I} for the $b_{-}$ branch, we find that Eq.~\eqref{DefPsirho}, in this case, is given by 
\begin{equation}
\Psi(\rho) = -\frac{\kappa \rho^{2}}{3\rho_{c}^{I}} + \frac{\kappa \rho}{3}\bigg( 1 - \frac{\rho}{\rho_{c}^{I}}\bigg)\frac{\gamma^2\rho/\rho_{c}^{I}}{(\gamma^2+1)(1+\sqrt{1-\rho/\rho_{c}^{I}})^2}
\label{Psirho1}
\end{equation}
or, using Eq.~\eqref{RmatterScalar},
\begin{equation}
\Psi(R_{T}) = -\frac{R_{T}^{2}}{12\rho_{c}^{I}\kappa} + \frac{R_{T}}{6}\bigg( 1 + \frac{R_{T}}{2\kappa\rho_{c}^{I}}\bigg)\frac{\gamma^2 R_{T}/(2\kappa\rho_{c}^{I})}{(\gamma^2+1)(1+\sqrt{1+R_{T}/(2\kappa\rho_{c}^{I}}))^2}\,.
\end{equation}
Therefore, in this case, Eq.~\eqref{solve3} provides
\begin{equation}
-\frac{\epsilon}{3}\left[\frac{1}{2} \varphi(R_{T}) - R_{T} \varphi^{\prime}(R_{T}) + 3 R_{T}^{2}  \varphi^{\prime\prime}(R_{T})\right]  =-\frac{R_{T}^{2}}{12\rho_{c}^{I}\kappa} + \frac{R_{T}}{6}\frac{\left( 1 + \frac{R_{T}}{2\kappa\rho_{c}^{I}}\right) \gamma^2 R_{T}/(2\kappa\rho_{c}^{I})}{(\gamma^2+1)\left(1+\sqrt{1+R_{T}/(2\kappa\rho_{c}^{I}})\right)^2} \,.
\label{DSolve1}
\end{equation}
Using \textit{Wolfram Mathematica}, we obtain the following solution
\begin{eqnarray}
\epsilon\varphi(R) &= &  c_1 R^{\frac{1}{6} \left(4-\sqrt{10}\right)}\epsilon + c_2 R^{\frac{1}{6} \left(\sqrt{10}+4\right)}\epsilon  + \frac{1}{90 \big(\gamma ^2+1\big) } \Bigg\{ 5 \left( - 72 \gamma ^2 \kappa  \rho_{c}^{I} +\frac{R^2}{\kappa  \rho_{c}^{I}}+54 \gamma ^2 R\right)
	\nonumber\\
&&+ \frac{18 \gamma ^2}{\sqrt{10}+10}\Bigg[3 \left(4 \left(5-4 \sqrt{10}\right) \kappa  \rho_{c}^{I} + \left(\sqrt{10}+10\right) R\right)  
\, _2F_1\left[-\frac{1}{2},\frac{1}{6} \left(-\sqrt{10}-4\right);\frac{1}{6} \left(2-\sqrt{10}\right);-\frac{R}{2 \kappa  \rho_{c}^{I}}\right] \nonumber\\
&&+ 3 \left(12 \left(2 \sqrt{10}+5\right) \kappa  \rho_{c}^{I}+\left(\sqrt{10}+10\right) R\right) 
\, _2F_1\left[-\frac{1}{2},\frac{1}{6} \left(\sqrt{10}-4\right);\frac{1}{6} \left(\sqrt{10}+2\right);-\frac{R}{2 \kappa  \rho_{c}^{I}}\right] \nonumber\\
&&+ (2 \kappa  \rho_{c}^{I}+R)\Bigg(9 \sqrt{10} \, _2F_1\left[\frac{1}{2},\frac{1}{6} \left(-\sqrt{10}-4\right);\frac{1}{6} \left(2-\sqrt{10}\right);-\frac{R}{2 \kappa  \rho_{c}^{I}}\right]  \nonumber\\
&&- \left(11 \sqrt{10}+20\right) \, _2F_1\left[\frac{1}{2},\frac{1}{6} \left(\sqrt{10}-4\right);\frac{1}{6} \left(\sqrt{10}+2\right);-\frac{R}{2 \kappa  \rho_{c}^{I}}\right]\Bigg) \Bigg] \Bigg\}\,,
\label{MathSol1}
\end{eqnarray}
where, again, we have dropped the subscript $T$ since, to $\epsilon$ order, $f(R)$ and $f(R_T)$ are the same.

As expected, the first two terms in Eq.~\eqref{MathSol1} correspond to Eq.~\eqref{Homo}, and the remaining terms correspond to the particular solution to Eq.~\eqref{DSolve1}. This solution contains hypergeometric functions\footnote{The Gaussian, or ordinary hypergeometric function, is a special function defined by the hypergeometric series
\begin{equation}\label{pow}
_2 F_1[a,b;c;z]=\sum_{n=0}^{\infty}\frac{(a)_n (b)_n}{(c)_n}\frac{z^n}{n!}\,
\end{equation}
on the disk $|z|<1$, where $(a)_n$, $(b)_n$, $(c)_n$ are the Pochhammer symbols defined as
\begin{equation}
(q)_n=\begin{cases} 1\,, \quad\quad n=0\,,\\ q(q+1)...(q+n-1)\,, \quad\quad n>0\,,\end{cases}
\end{equation}
and by analytic continuation elsewhere. $_2F_1[a,b;c;z]$ has a branch cut discontinuity in the complex plane running from $1$ to $\infty$. The hypergeometric series is convergent for arbitrary $a$, $b$ and $c$ for real $-1<z<1$ and for $z\pm 1$ if $c > a + b$.}, denoted by $_2F_1$, and its domain is $- 2\kappa \rho_{c}^{I} \leq R$. Along with the fact that it only makes physical sense that $R \leq 0$, we have that our physical solution is defined in the range
\begin{equation}
- 2\kappa \rho_{c}^{I} \leq R \leq 0\,.
\label{intervalHyper1}
\end{equation}

Thus, a metric $f(R)$ action was found which, when treated as an effective action, leads to the modified Friedmann equation \eqref{H2I2}, provided by the $b_{-}$ branch of mLQC-I. Setting $c_{1}=c_{2}=0$, without loss of generality, the effective action is given by
\begin{eqnarray}
f(R) &=& R + \frac{1}{90 \big(\gamma ^2+1\big) } \Bigg\{ 5 \left( - 72 \gamma ^2 \kappa  \rho_{c}^{I} + \frac{R^2}{\kappa  \rho_{c}^{I}}+54 \gamma ^2 R\right)+ \frac{18 \gamma ^2}{\sqrt{10}+10}\Bigg[3 \bigg(4 \left(5-4 \sqrt{10}\right) \kappa  \rho_{c}^{I}  
	\nonumber\\
&& + \left(\sqrt{10}+10\right) R\bigg) \, _2F_1\left[-\frac{1}{2},\frac{1}{6} \left(-\sqrt{10}-4\right);\frac{1}{6} \left(2-\sqrt{10}\right);-\frac{R}{2 \kappa  \rho_{c}^{I}}\right] 
	\nonumber\\
&&+ 3 \left(12 \left(2 \sqrt{10}+5\right) \kappa  \rho_{c}^{I}+\left(\sqrt{10}+10\right) R\right) 
\, _2F_1\left[-\frac{1}{2},\frac{1}{6} \left(\sqrt{10}-4\right);\frac{1}{6} \left(\sqrt{10}+2\right);-\frac{R}{2 \kappa  \rho_{c}^{I}}\right] 
	\nonumber\\
&&+ (2 \kappa  \rho_{c}^{I}+R)\bigg(9 \sqrt{10} \; _2F_1\left[\frac{1}{2},\frac{1}{6} \left(-\sqrt{10}-4\right);\frac{1}{6} \left(2-\sqrt{10}\right);-\frac{R}{2 \kappa  \rho_{c}^{I}}\right]  \nonumber\\
&& - \left(11 \sqrt{10}+20\right) \, _2F_1\left[\frac{1}{2},\frac{1}{6} \left(\sqrt{10}-4\right);\frac{1}{6} \left(\sqrt{10}+2\right);-\frac{R}{2 \kappa  \rho_{c}^{I}}\right]\bigg) \Bigg] \Bigg\} \,.
\label{fRbminus}
\end{eqnarray}
Using \textit{Wolfram Mathematica}, a Taylor expansion of Eq.~\eqref{fRbminus}, about $R=0$ and up to third order, was computed and is given by
\begin{equation}
f(R) = R + \epsilon \varphi(R) = R + \frac{(4+3\gamma^2)R^{2}}{72(1+\gamma^{2})\kappa\rho_{c}^{I}} + \frac{\gamma^{2}R^{3}}{992 (1+\gamma^{2}) (\kappa\rho_{c}^{I})^{2}} + ...\,,
\label{TaylormLQCIbminusW1}
\end{equation}
where the critical density for this model is given by
\begin{equation}
\rho_{c}^{I} \equiv \frac{\sqrt{3}}{128 \pi^{2} \gamma^{3} (1+\gamma^{2})}\,.
\label{rhocIGeoUnit}
\end{equation}

\begin{figure}[htb!]
\centering%
\subfigure[~Deviation, from GR, of the effective Lagrangian: a comparison between the $b_{-}$ branch of mLQC-I and LQC.] {\label{bMinusPlot1}
\includegraphics[width=.49\textwidth]{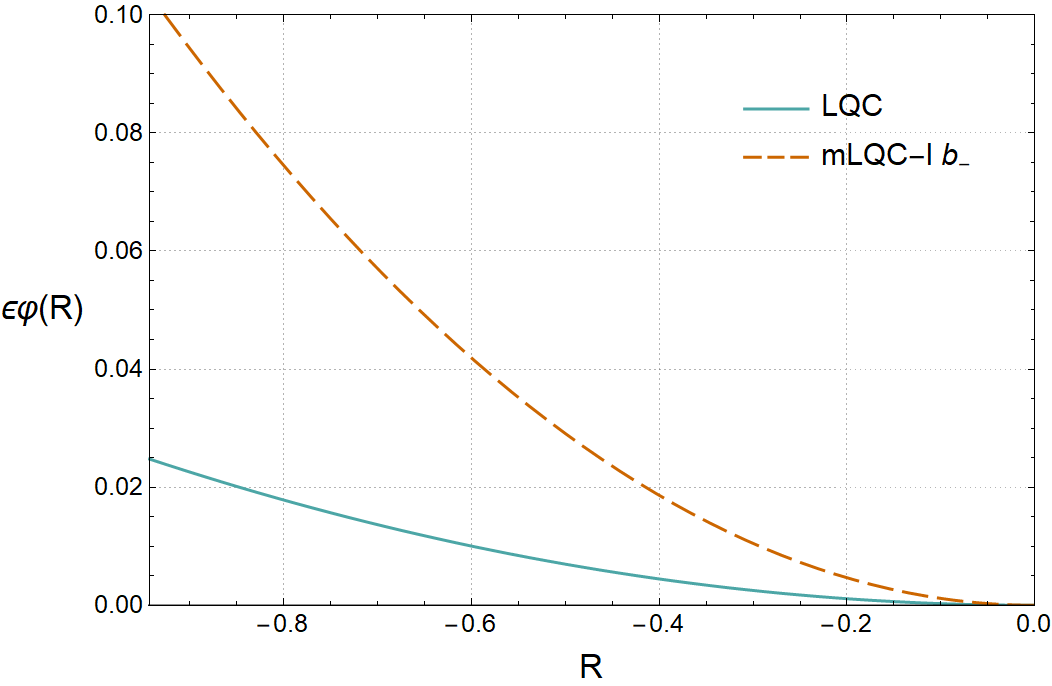} }
\subfigure[~Effective $f(R)$ function: a comparison between the $b_{-}$ branch of mLQC-I, LQC and GR.] {\label{bMinusPlot2}
\includegraphics[width=.48\textwidth]{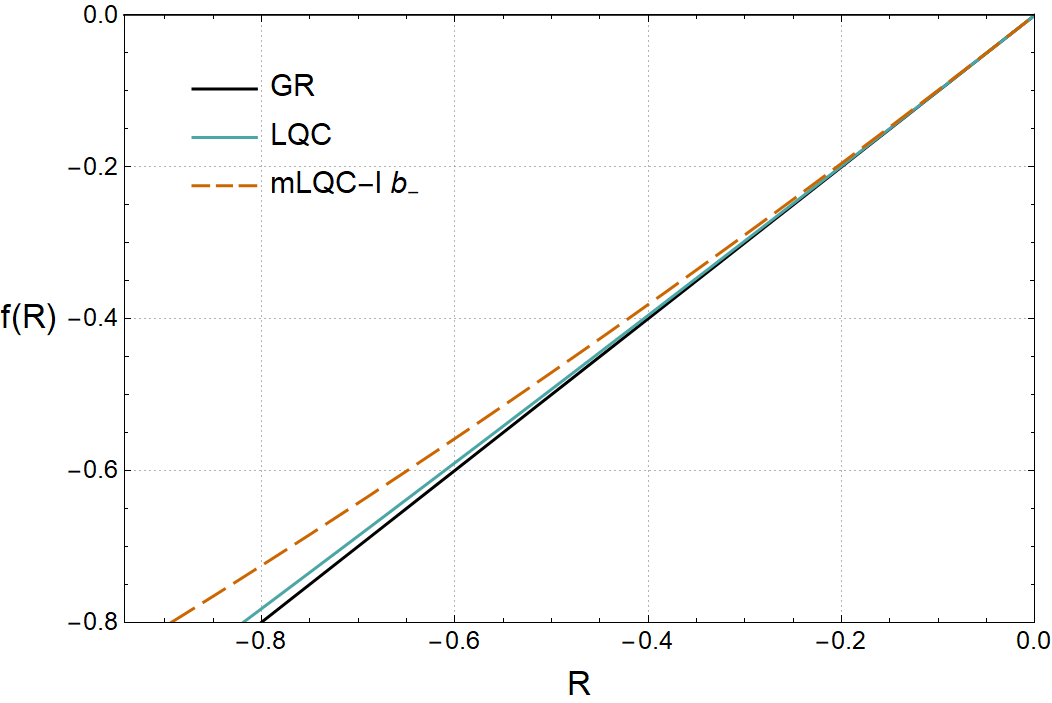} }
\caption{The plot~\ref{bMinusPlot1} presents the comparison between Eq.~\eqref{MathSol1}, for the $b_{-}$ branch of mLQC-I (dashed dark orange line), and Eq.~\eqref{deviationLQC}, for LQC (solid blue line), with $c_1=c_2=0$, for $- 2\kappa \rho_{c}^{I} \leq R \leq 0$. Although the deviation is higher for the case of the $b_{-}$ branch of this modified model, both of them are dominated by the quadratic term.
The plot~\ref{bMinusPlot2} illustrates the differences between the Lagrangian density of GR (solid black line), with $\Lambda = 0$, and the effective Lagrangian densities given by Eq.~\eqref{fRLQC}, in the case of LQC (solid blue line) and Eq.~\eqref{fRbminus}, for the $b_{-}$ branch of mLQC-I (dashed dark orange line), for the interval $- 2\kappa \rho_{c}^{I} \leq R \leq 0$.}
\label{fig1}
\end{figure}

In order to better convey the magnitude of the deviation from GR in this case, we provide two different plots, in Fig.~\ref{fig1}. In Fig.~\ref{bMinusPlot1} we present the deviation of the effective Lagrangian density from the one we find in GR, as a function of $R$. In this plot, we are able to compare the deviation in the case of LQC, given by Eq.~\eqref{deviationLQC}, with that of mLQC-I, in the case of $b_{-}$ branch. Notice that we are plotting Eq.~\eqref{MathSol1}, with $c_{1}=c_{2}=0$, not the corresponding series expansion. In Fig.~\ref{bMinusPlot2}, we compare the effective $f(R)$ functions of LQC and the $b_{-}$ branch of mLQC-I, given by Eq.~\eqref{fRLQC} and Eq.~\eqref{fRbminus}, respectively, with that of GR, with $\Lambda = 0$. 

%%%%%%%%%%%%%%%%%%%%%%%%%%%%%%%%%%%%%%
\subsubsection{Effective action for the $b_{+}$ branch}

Considering the $b_{+}$ branch, with the modified Friedmann equation \eqref{H2I}, we find that Eq.~\eqref{DefPsirho} is given by 
\begin{equation}
\Psi(\rho) = -\frac{\kappa  \rho }{3} + \frac{1}{3} \alpha  \kappa  \rho_{\Lambda}  \left(1-\frac{\rho }{\rho_{c}^{I}}\right) \left[1 + \frac{\rho  \left(1-2 \gamma ^2+\sqrt{1 - \rho/\rho_{c}^{I}}\right)}{4 \gamma^2 \rho_{c}^{I} \left(1 + \sqrt{1-\rho/\rho_{c}^{I} }\right)}\right] \,,
\label{Psirho2}
\end{equation}
or, using Eq.~\eqref{RmatterScalar},
\begin{equation}
\Psi(R_{T}) =\frac{R_{T}}{6} + \frac{1}{3} \alpha  \kappa  \rho_{\Lambda} \left(1 + \frac{R_{T}}{2 \kappa  \rho_{c}^{I}}\right) \left[1-\frac{R_{T} \left(1 -2 \gamma ^2+\sqrt{1 + R_{T}/(2 \kappa  \rho_{c}^{I})}\right)}{8 \gamma ^2 \kappa  \rho_{c}^{I} \left(1 + \sqrt{1 + R_{T}/(2 \kappa  \rho_{c}^{I})}\right)}\right]\,.
\end{equation}
Therefore, in this case Eq.~\eqref{solve3} becomes
\begin{equation}
-\frac{\epsilon}{3}\left[\frac{1}{2} \varphi_{T} - R_{T} \varphi^{\prime}_{T} + 3 R_{T}^{2}  \varphi^{\prime\prime}_{T}\right]  = \frac{R_{T}}{6} + \frac{1}{3}\alpha  \kappa  \rho_{\Lambda} \left(1 + \frac{R_{T}}{2 \kappa  \rho_{c}^{I}}\right)  \left[1-\frac{R_{T} \left(1 -2 \gamma ^2+\sqrt{1 + R_{T}/(2 \kappa  \rho_{c}^{I})}\right)}{8 \gamma^2 \kappa  \rho_{c}^{I} \left(1 + \sqrt{1 + R_{T}/(2 \kappa  \rho_{c}^{I}) }\right)}\right] \,,
\label{DSolve2}
\end{equation}
where $\varphi_{T} = \varphi(R_{T})$. As before, the solution to Eq.~\eqref{DSolve2} was obtained using \textit{Wolfram Mathematica}, and is given by
\begin{eqnarray}
\epsilon\varphi(R) &=&  c_1  R^{\frac{1}{6} \left(4-\sqrt{10}\right)}\epsilon  + c_2 R^{\frac{1}{6} \left(\sqrt{10}+4\right)}\epsilon -\alpha  \kappa  \rho_\Lambda + R +\frac{\alpha  \rho_\Lambda  R \left(18 \left(2 \gamma ^2-1\right) \kappa  \rho_{c}^{I} +R\right)}{72 \gamma ^2 \kappa (\rho_{c}^{I})^2}
	\nonumber\\
&&+\frac{\alpha \rho_\Lambda}{20 \rho_{c}^{I}} \Bigg\{-3 \left(R-2 \left(\sqrt{10}-2\right) \kappa  \rho_{c}^{I}\right) \, _2F_1\left[-\frac{1}{2},\frac{1}{6} \left(-\sqrt{10}-4\right);\frac{1}{6} \left(2-\sqrt{10}\right);-\frac{R}{2 \kappa  \rho_{c}^{I}}\right] 
	\nonumber\\
&& - 3 \left(2 \left(\sqrt{10}+2\right) \kappa  \rho_{c}^{I}+R\right) \, _2F_1\left[-\frac{1}{2},\frac{1}{6} \left(\sqrt{10}-4\right);\frac{1}{6} \left(\sqrt{10}+2\right);-\frac{R}{2 \kappa  \rho_{c}^{I}}\right]
	\nonumber\\
&&+(2 \kappa  \rho_{c}^{I}+R) \Bigg(-\left(\sqrt{10}-1\right) \, _2F_1\left[\frac{1}{2},\frac{1}{6} \left(-\sqrt{10}-4\right);\frac{1}{6} \left(2-\sqrt{10}\right);-\frac{R}{2 \kappa  \rho_{c}^{I}}\right]
	\nonumber\\
&& + \left(\sqrt{10}+1\right) \, _2F_1\left[\frac{1}{2},\frac{1}{6} \left(\sqrt{10}-4\right);\frac{1}{6} \left(\sqrt{10}+2\right);-\frac{R}{2 \kappa  \rho_{c}^{I}}\right]\Bigg)\Bigg\}\,,
\label{MathSol2}
\end{eqnarray}
where we have dropped the subscript $T$ since, to $\epsilon$ order, $f(R)$ and $f(R_{T})$ are the same. Once again, the first two terms correspond to Eq.~\eqref{Homo}, and the remaning ones correspond to the particular solution of Eq.~\eqref{DSolve2}. As in the previous branch, the solution contains hypergeometric functions and its domain is given by $- 2 \kappa \rho_{c}^{I} \leq R$. As such, since it only makes physical sense that $R\leq 0$, Eq.~\eqref{MathSol2} is defined, as a physical solution, for the range
\begin{equation}
- 2 \kappa \rho_{c}^{I} \leq R \leq 0\,.
\label{intervalHyper2}
\end{equation}
Taking the same argument as in the previous calculations, we can set $c_{1}=c_{2}=0$ and, as a consequence, the effective action is given by
\begin{eqnarray}
f(R) & = & -\alpha  \kappa  \rho_\Lambda + 2 R +\frac{\alpha  \rho_\Lambda  R \left(18 \left(2 \gamma ^2-1\right) \kappa  \rho_{c}^{I} +R\right)}{72 \gamma ^2 \kappa (\rho_{c}^{I})^2} 
	\nonumber \\
&&
+\frac{\alpha \rho_\Lambda}{20 \rho_{c}^{I}} \Bigg\{-3 \left[R-2 \left(\sqrt{10}-2\right) \kappa  \rho_{c}^{I}\right]
	 \, _2F_1\left[-\frac{1}{2},\frac{1}{6} \left(-\sqrt{10}-4\right);\frac{1}{6} \left(2-\sqrt{10}\right);-\frac{R}{2 \kappa  \rho_{c}^{I}}\right] 
\nonumber\\
&&
- 3 \left(2 \left(\sqrt{10}+2\right) \kappa  \rho_{c}^{I}+R\right)
	 \, _2F_1\left[-\frac{1}{2},\frac{1}{6} \left(\sqrt{10}-4\right);\frac{1}{6} \left(\sqrt{10}+2\right);-\frac{R}{2 \kappa  \rho_{c}^{I}}\right]
\nonumber\\
&&
+ (2 \kappa  \rho_{c}^{I}+R)
	 \Bigg(-\left(\sqrt{10}-1\right) \, _2F_1\left[\frac{1}{2},\frac{1}{6} \left(-\sqrt{10}-4\right);\frac{1}{6} \left(2-\sqrt{10}\right);-\frac{R}{2 \kappa  \rho_{c}^{I}}\right]
\nonumber\\
&&+\left(\sqrt{10}+1\right) \, _2F_1\left[\frac{1}{2},\frac{1}{6} \left(\sqrt{10}-4\right);\frac{1}{6} \left(\sqrt{10}+2\right);-\frac{R}{2 \kappa  \rho_{c}^{I}}\right]\Bigg)\Bigg\}\,.
\label{fRbplus}
\end{eqnarray}
A Taylor expansion of Eq.~\eqref{fRbplus} up to third order, about $R=0$, was computed using \textit{Wolfram Mathematica}, resulting in
\begin{equation}
f(R) = -2 \alpha  \kappa  \rho_\Lambda + \left(\frac{\alpha  \left(5 \gamma ^2-1\right) \rho_\Lambda }{4 \gamma^2 \rho_{c}^{I}}+2\right) R +\frac{\alpha  \left(4-3 \gamma^2\right) \rho_\Lambda  R^2}{288 \gamma ^2 \kappa (\rho_{c}^{I})^2}+\frac{\alpha  \rho_\Lambda  R^3}{3968 \kappa^2 (\rho_{c}^{I})^3}+... \,\, .
\label{TaylormLQCIbplusW1}
\end{equation}
We observe that, for this particular branch, $f(0) \ne 0$. We recall that $\rho_{\Lambda}$ was interpreted as the energy density of an emergent positive cosmological constant in Ref.~\cite{Li:2018opr}. 

\begin{figure}[htb!]
\centering%
\subfigure[~Deviation, from GR, of the effective Lagrangian: a comparison between the $b_{+}$ branch of mLQC-I and LQC.] {\label{bPlusPlot1}
\includegraphics[width=.49\textwidth]{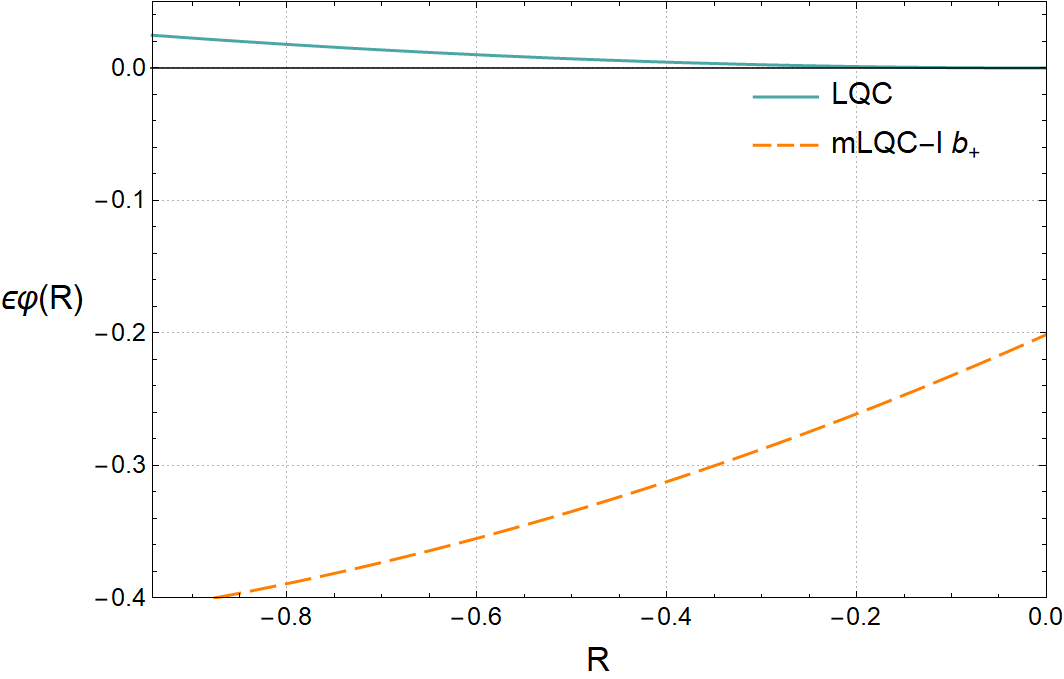} }
\subfigure[~Effective $f(R)$ function: a comparison between the $b_{+}$ branch of mLQC-I, LQC and GR.] {\label{bPlusPlot2}
\includegraphics[width=.48\textwidth]{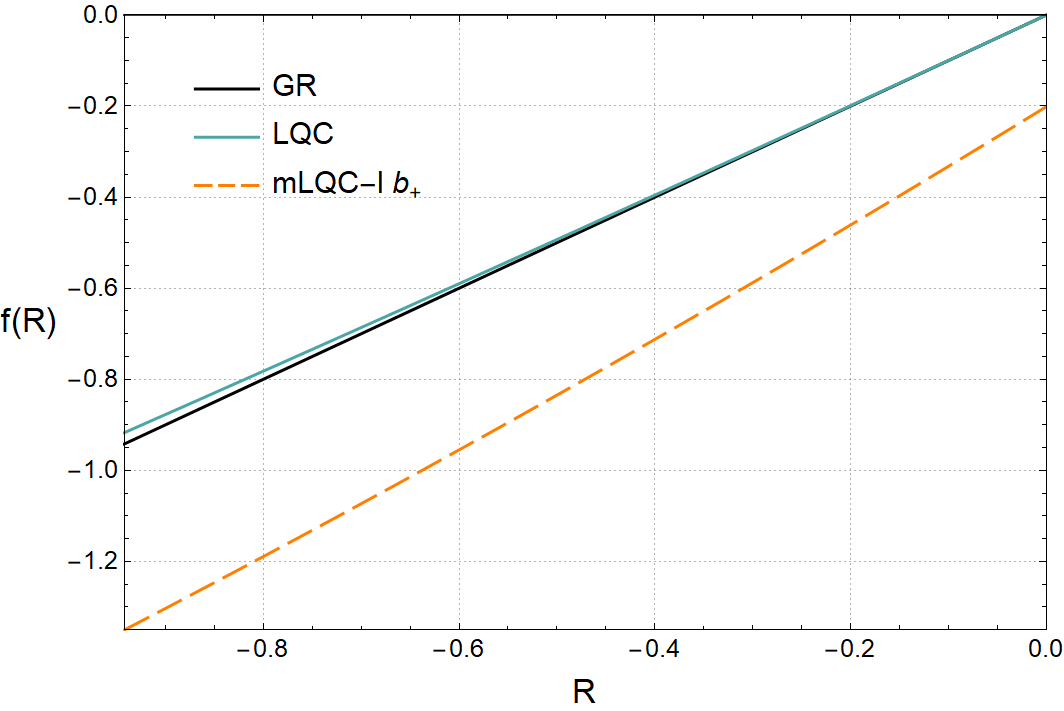} }
\caption{The plot~\ref{bPlusPlot1} presents the comparison between Eq.~\eqref{MathSol2}, for the $b_{+}$ branch of mLQC-I (dashed light orange line), and Eq.~\eqref{deviationLQC}, for LQC (solid blue line), with $c_1=c_2=0$, for $- 2\kappa \rho_{c}^{I} \leq R \leq 0$. Contrary to the previous case, for this branch the correction is dominated by the linear term. As a result, since $R<0$, the deviation is predominantly negative.
The plot~\ref{bPlusPlot2} illustrates the differences between the Lagrangian density of GR (solid black line), with $\Lambda=0$, and the effective Lagrangian densities given by Eq.~\eqref{fRLQC}, in the case of LQC (solid blue line), and Eq.~\eqref{fRbminus}, for the $b_{+}$ branch of mLQC-I (dashed light orange line), for the interval $- 2\kappa \rho_{c}^{I} \leq R \leq 0$.}
\label{fig2}
\end{figure}

Thus, a metric $f(R)$ action was also found for the $b_{+}$ branch of mLQC-I which, when treated as an effective action, leads to the modified Friedmann Eq.~\eqref{H2I2}. Finally, using the same definitions we used before, to provide plots for the $b_{-}$ branch, we also provide the same kind of plots for this branch, in Fig.~\ref{fig2}. In Fig.~\ref{bPlusPlot1}, we can observe that, for mLQC-II, $\epsilon\varphi(R)$ is an increasing function, contrary to all the other models we are considering.

%%%%%%%%%%%%%%%%%%%%%%%%%%%%%%%%%%%%%%%%%%%%%%%%%%%%%%%%%%%%%%%%%
\subsection{Effective action for mLQC-II}
%%%%%%%%%%%%%%%%%%%%%%%%%%%%%%%%%%%%%%%%%%%%%%%%%%%%%%%%%%%%%%%%%

We now consider the mLQC-II model, in which the modified Friedmann equation is given by Eq.~\eqref{H2II}. In this case, Eq.~\eqref{DefPsirho} is given by
\begin{equation}
\Psi(\rho) = \frac{2 \kappa  \rho  \left(1- \rho/\rho_{c}^{II}\right) \left(1 + 4 \left(\gamma ^2+1\right) \gamma ^2 \rho/\rho_{c}^{II} \right)}{3 \left(1 + 2 \gamma ^2 \rho / \rho_{c}^{II} +\sqrt{1 + 4 \left(\gamma ^2+1\right) \gamma ^2 \rho/\rho_{c}^{II} }\right)}-\frac{\kappa  \rho }{3}\,,
\label{Psirho3}
\end{equation}
or, using Eq.~\eqref{RmatterScalar},
\begin{equation}
\Psi(R_{T}) =\frac{R_{T}}{6}-\frac{R_{T} \left(1 + R_{T}/(2 \kappa  \rho_{c}^{II}) \right) \left(1- 2 \gamma ^2 \left(\gamma ^2+1\right) R_{T}/(\kappa \rho_{c}^{II})\right)}{3 \left(1- \gamma ^2 R_{T}/(\kappa  \rho_{c}^{II}) +\sqrt{1- 2 \gamma ^2 \left(\gamma ^2+1\right) R_{T}/(\kappa \rho_{c}^{II})}\right)}\,.
\end{equation}
Therefore, in this case Eq.~\eqref{solve3} is given by
\begin{equation}
-\frac{\epsilon}{3}\left[\frac{1}{2} \varphi(R_{T}) - R_{T} \varphi^{\prime}(R_{T}) + 3 R_{T}^{2}  \varphi^{\prime\prime}(R_{T})\right]  = \frac{R_{T}}{6}-\frac{R_{T} \left(1 + \frac{R_{T}}{2 \kappa  \rho_{c}^{II}}\right) \left(1-\frac{2 \gamma ^2 \left(\gamma ^2+1\right) R_{T}}{\kappa \rho_{c}^{II}}\right)}{3 \left(1-\frac{\gamma ^2 R_{T}}{\kappa  \rho_{c}^{II}}+\sqrt{1-\frac{2 \gamma ^2 \left(\gamma ^2+1\right) R_{T}}{\kappa \rho_{c}^{II}}}\right)}\,.
\label{DSolve3}
\end{equation}
The solution to Eq.~\eqref{DSolve3} was obtained using \textit{Wolfram Mathematica}, and leads to 
\begin{eqnarray}
\epsilon\varphi(R) &= & c_1 R^{\frac{1}{6} \left(4-\sqrt{10}\right)}\epsilon + c_2  R^{\frac{1}{6} \left(\sqrt{10}+4\right)}\epsilon  + \frac{1}{90 \gamma^4 }\Bigg\{ 270 \left(\gamma ^4+\gamma ^2\right) R+\frac{20 \left(\gamma ^6+\gamma ^4\right) R^2}{\kappa \rho_{c}^{II}} 
+ 90 \kappa  \rho_{c}^{II} 
	\nonumber\\
&&+ 9 \Bigg[3 \left(\left(\sqrt{10}-2\right) \kappa  \rho_{c}^{II}+2 \left(\gamma ^4+\gamma ^2\right) R\right)
	 \, _2F_1\left[-\frac{1}{2},\frac{1}{6} \left(-\sqrt{10}-4\right);\frac{1}{6} \left(2-\sqrt{10}\right);\frac{2 R \gamma ^2 \left(\gamma ^2+1\right)}{\kappa  \rho_{c}^{II}}\right]
	\nonumber\\
&&+ \left(6 \left(\gamma ^4+\gamma ^2\right) R-3 \left(\sqrt{10}+2\right) \kappa  \rho_{c}^{II}\right) 
	\, _2F_1\left[-\frac{1}{2},\frac{1}{6} \left(\sqrt{10}-4\right);\frac{1}{6} \left(\sqrt{10}+2\right);\frac{2 R \gamma ^2 \left(\gamma ^2+1\right)}{\kappa  \rho_{c}^{II}}\right]
	\nonumber\\
&&+ \left(2 \left(\gamma ^4+\gamma ^2\right) R-\kappa \rho_{c}^{II}\right)
	\Bigg(\left(\sqrt{10}-1\right) \, _2F_1\left[\frac{1}{2},\frac{1}{6} \left(-\sqrt{10}-4\right);\frac{1}{6} \left(2-\sqrt{10}\right);\frac{2 R \gamma ^2 \left(\gamma ^2+1\right)}{\kappa  \rho_{c}^{II}}\right]
	\nonumber\\
&&-\left(\sqrt{10}+1\right) \, _2F_1\left[\frac{1}{2},\frac{1}{6} \left(\sqrt{10}-4\right);\frac{1}{6} \left(\sqrt{10}+2\right);\frac{2 R \gamma ^2 \left(\gamma ^2+1\right)}{\kappa  \rho_{c}^{II}}\right]\Bigg)\Bigg]\Bigg\}\,,
\label{MathSol3}
\end{eqnarray}
where we have dropped the subscript $T$ since, to order $\epsilon$, $f(R)$ and $f(R_{T})$ are the same. As usual, the first two terms correspond to Eq.~\eqref{Homo} and the remaning terms correspond to the particular solution to Eq.~\eqref{DSolve3}. For this model, the solution also contains hypergeometric functions but with a different argument and its domain is given by $R \leq \kappa \rho_{c}^{II} /\left(2 \gamma^{2}(\gamma^2 + 1)\right)$. As such, in this case the solution only has physical meaning for the interval
\begin{equation}
R \leq 0\,.
\label{intervalHyper3}
\end{equation}

Thus, a metric $f(R)$ action was found which, when treated as an effective action, leads to the modified Friedmann Eq.~\eqref{H2II}, provided by mLQC-II. Setting $c_{1}=c_{2}=0$ this effective action is given by
\begin{eqnarray}
f(R) & = & R + \frac{1}{90 \gamma^4 }\Bigg\{ 270 \left(\gamma ^4+\gamma ^2\right) R+\frac{20 \left(\gamma ^6+\gamma ^4\right) R^2}{\kappa \rho_{c}^{II}}+90 \kappa  \rho_{c}^{II}+9 \Bigg[3\Big (\left(\sqrt{10}-2\right) \kappa  \rho_{c}^{II}
	\nonumber\\
&& +2 \left(\gamma ^4+\gamma ^2\right) R\Big) \, _2F_1\left[-\frac{1}{2},\frac{1}{6} \left(-\sqrt{10}-4\right);\frac{1}{6} \left(2-\sqrt{10}\right);\frac{2 R \gamma ^2 \left(\gamma ^2+1\right)}{\kappa  \rho_{c}^{II}}\right]
	\nonumber\\
&& + \left(6 \left(\gamma ^4+\gamma ^2\right) R-3 \left(\sqrt{10}+2\right) \kappa  \rho_{c}^{II}\right)
	 \, _2F_1\left[-\frac{1}{2},\frac{1}{6} \left(\sqrt{10}-4\right);\frac{1}{6} \left(\sqrt{10}+2\right);\frac{2 R \gamma ^2 \left(\gamma ^2+1\right)}{\kappa  \rho_{c}^{II}}\right]
\nonumber \\
&&	 
	 +\left(2 \left(\gamma ^4+\gamma ^2\right) R-\kappa \rho_{c}^{II}\right)
	\Bigg(\left(\sqrt{10}-1\right)\, _2F_1\left[\frac{1}{2},\frac{1}{6} \left(-\sqrt{10}-4\right);\frac{1}{6} \left(2-\sqrt{10}\right);\frac{2 R \gamma ^2 \left(\gamma^2+1\right)}{\kappa  \rho_{c}^{II}}\right]
	\nonumber\\
&&-\left(\sqrt{10}+1\right) \, _2F_1\left[\frac{1}{2},\frac{1}{6} \left(\sqrt{10}-4\right);\frac{1}{6} \left(\sqrt{10}+2\right);\frac{2 R \gamma ^2 \left(\gamma ^2+1\right)}{\kappa  \rho_{c}^{II}}\right]\Bigg)\Bigg]\Bigg\} \,.\label{fRII}
\end{eqnarray}
Using \textit{Wolfram Mathematica}, a Taylor expansion of Eq.~\eqref{fRII}, about $R=0$ and up to third order, was computed and is given by
\begin{equation}
f(R) = R + \frac{\left(-3 \gamma^4 - 2 \gamma^2 + 1\right) R^2}{18 \kappa  \rho_{c}^{II}} -\frac{\gamma ^2 \left(\gamma ^2+1\right)^3  R^3}{62 (\kappa \rho_{c}^{II})^2} + ... \,\, ,
\label{TaylormLQCIIW1}
\end{equation}
where the critical density for this model is
\begin{equation}
\rho_{c}^{II} \equiv \frac{\sqrt{3}(\gamma^{2} + 1)}{8\pi^{2} \gamma^{3} }\,.
\label{rhocIIGeoUnit}
\end{equation}

\begin{figure}[htb!]
\centering%
\subfigure[~Deviation, from GR, of the effective Lagrangian: a comparison between mLQC-II and LQC.] {\label{IIPlot1}
\includegraphics[width=.4855\textwidth]{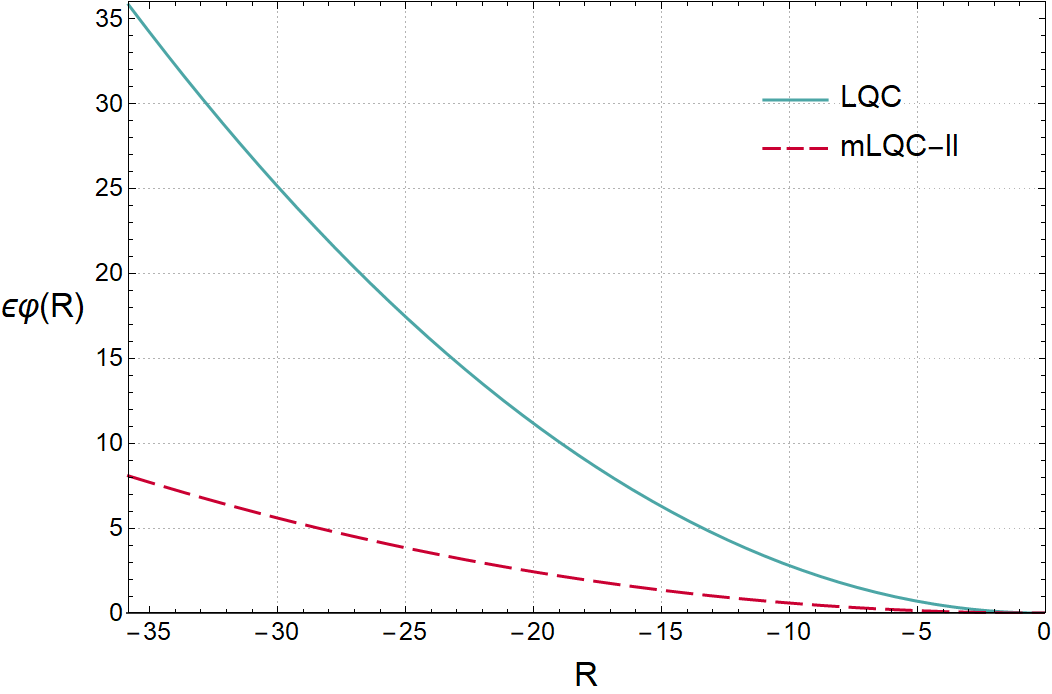} }
\subfigure[~Effective $f(R)$ function: a comparison between mLQC-II, LQC and GR.] {\label{IIPlot2}
\includegraphics[width=.48\textwidth]{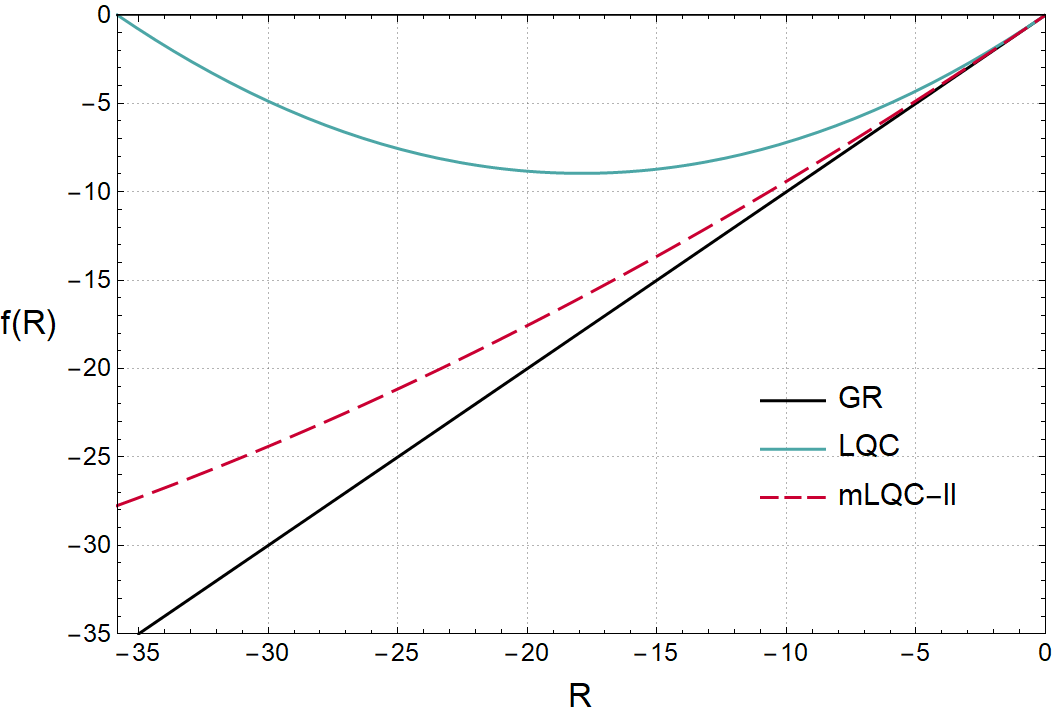} }
\caption{The plot~\ref{IIPlot1} presents the comparison between Eq.~\eqref{MathSol3}, for the case of mLQC-II (dashed pink line), and Eq.~\eqref{deviationLQC}, for LQC (solid blue line), with $c_1=c_2=0$, for $-18\kappa\rho_{c}\leq R \leq 0$. Both of the corrections are dominated by the quadratic term.
The plot~\ref{IIPlot2} illustrates the differences between the Lagrangian density of GR (solid black line), with $\Lambda =0$, and the effective Lagrangian densities, in the case of LQC (solid blue line), and Eq.~\eqref{fRII}, for mLQC-II (dashed pink line), for the interval $-18\kappa\rho_{c}\leq R \leq 0$.}
\label{fig3}
\end{figure}

This solution for the mLQC-II model is depicted in Fig.~\ref{fig3}. In Fig.~\ref{IIPlot1}, we present the deviation of the effective Lagrangian density from the general relativistic counterpart, and compare them to the the deviation in the case of LQC, given by Eq.~\eqref{deviationLQC}, with that of mLQC-II, given by Eq.~\eqref{MathSol1} with $c_{1}=c_{2}=0$. In Fig.~\ref{IIPlot2}, we compare the effective $f(R)$ functions of LQC and mLQC-II, given by Eq.~\eqref{fRLQC} and Eq.~\eqref{fRII}, respectively, with that of GR, with $\Lambda = 0$. 

\begin{figure}[htb!]
\centering%
\subfigure[~Deviation, from GR, of the effective Lagrangian: a comparison between all models.] {\label{AllPlot1}
\includegraphics[width=.485\textwidth]{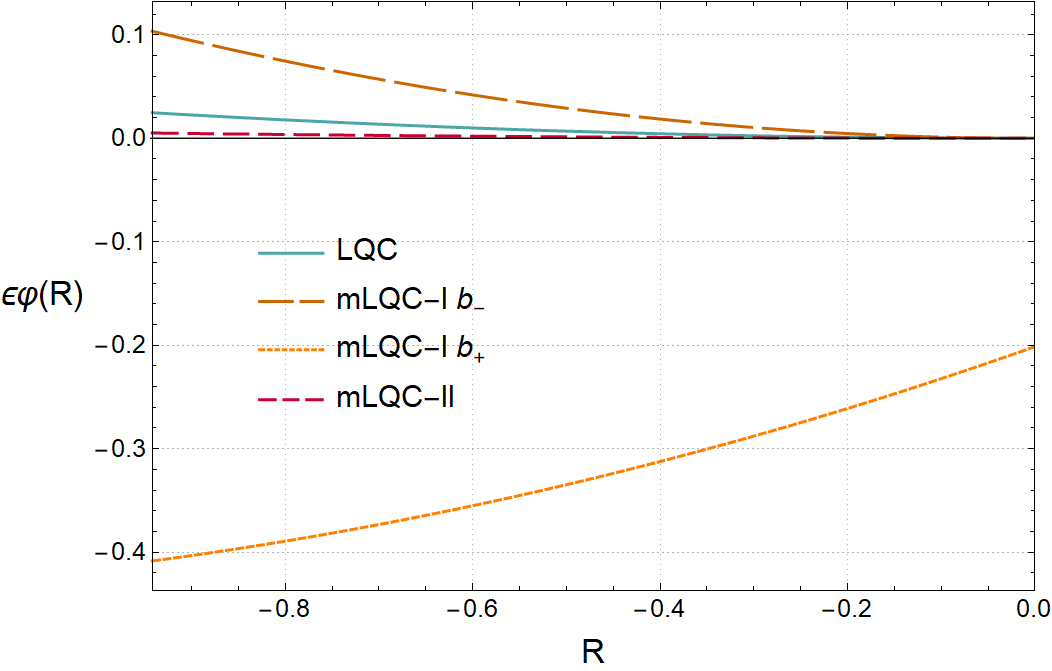} }
\subfigure[~Effective $f(R)$ function: a comparison between all models.] {\label{AllPlot2}
\includegraphics[width=.48\textwidth]{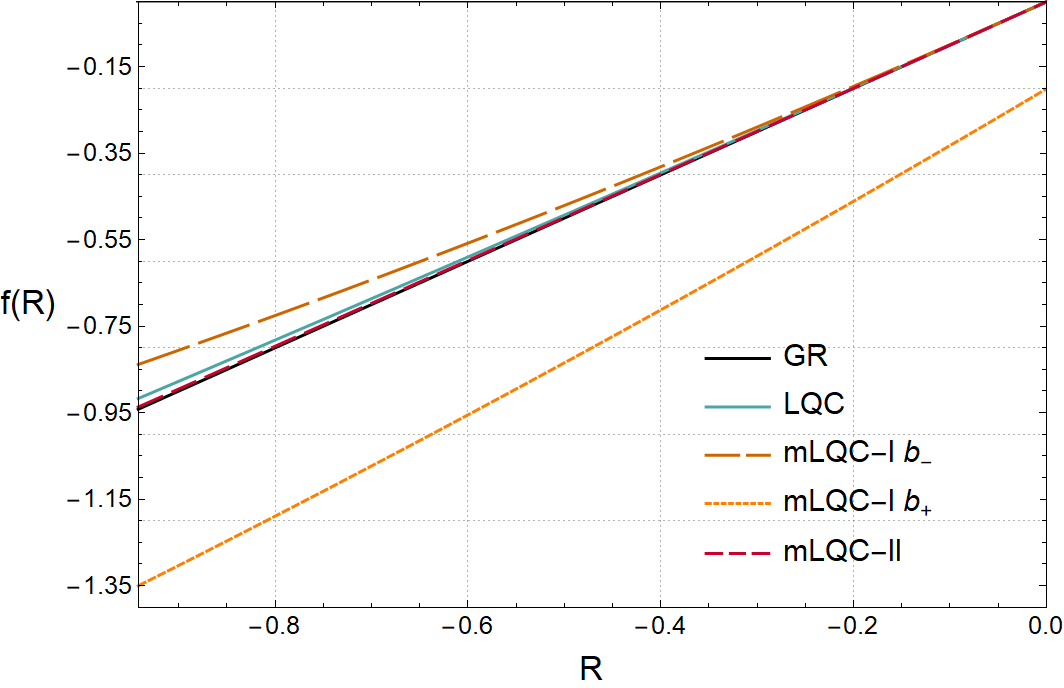} }
\caption{The plot~\ref{AllPlot1} contains Eq.~\eqref{deviationLQC}, for LQC (solid blue line), Eq.~\eqref{MathSol1}, for the $b_{-}$ branch of mLQC-I (dashed dark orange line), Eq.~\eqref{MathSol2}, for the $b_{+}$ branch of mLQC-I (dashed light orange line)  and Eq.~\eqref{MathSol3}, for mLQC-II (dashed pink line), with $c_1=c_2=0$, for $- 2\kappa \rho_{c}^{I} \leq R \leq 0$.
The plot~\ref{AllPlot2} presents the effective $f(R)$ functions for all models, including GR (solid black line), with $\Lambda =0$. These are Eq.~\eqref{fRLQC}, for LQC (solid blue line), Eq.~\eqref{fRbminus}, for the $b_{-}$ branch of mLQC-I (dashed dark orange line), Eq.~\eqref{fRbplus}, for $b_{+}$ branch of mLQC-I (dashed light orange line) and Eq.~\eqref{fRII}, for mLQC-II (dashed pink line), for $- 2\kappa \rho_{c}^{I} \leq R \leq 0$.}
\label{fig4}
\end{figure}

Finally, we also present in Fig.~\ref{fig4} two plots with all the solutions. In Fig.~\ref{AllPlot1} we compare all the deviations with respect to the Lagrangian density of GR. In Fig.~\ref{AllPlot2} all the effective metric $f(R)$ functions are displayed. With the exception of the $b_{+}$ branch of mLQC-I, all deviations from GR are dominated by the quadratic term and the remaning terms become more subdominat, with increasing order, as is the case in LQC. In the case of the $b_{+}$ branch, the dominant term is linear, due to the presence of the constant on the right-hand side of Eq.~\eqref{DSolve2}, which is not the case in any of the other equations we have solved.

%%%%%%%%%%%%%%%%%%%%%%%%%%%%%%%%%%%%%%%%%%%%%%%%%%%%%%%%%%%%%%%%%
\section{Conclusions}\label{Sec:Conclusions}
%%%%%%%%%%%%%%%%%%%%%%%%%%%%%%%%%%%%%%%%%%%%%%%%%%%%%%%%%%%%%%%%%

In this work, we have addressed the initial singularity problem that is present in the $\Lambda$CDM model, in which the Universe emerges from a single point with infinite density, and considered a possible resolution to this question, as proposed by LQC, in which the Big Bang is replaced by a quantum bounce. Through an effective Hamiltonian description, this bounce is manifested by an effective Friedmann equation that is similar to the one found in GR, but with a modified source, given by Eq.~\eqref{LQC}. It is clear from this equation that the energy density reaches a finite maximum value, a critical density, contrary to the Big Bang scenario. In this regard, LQC proposes a scenario in which the Universe undergoes a collapse to an expansion through a bounce. 

Since this result comes from a different field in physics, it is relevant to ask if it is possible to replicate it in the framework of GR and its modifications. In this context, in order to obtain the modified Friedmann equation~\eqref{LQC}, we considered the class of metric $f(R)$ gravity, where the Ricci scalar in the EH action, is substituted by a general function of $R$. 
Using this particular modification of GR, an effective action that leads to Eq.~\eqref{LQC} was determined in Ref.~\cite{Sotiriou:2008ya}, assuming matter as a scalar field. Motivated by this approach, we considered two modifications of standard LQC, which also yield a quantum bounce that occurs for a critical density, specific to each model, which we denoted by mLQC-I and mLQC-II, as outlined in the Introduction. These modifications came to be formulated as a result of a departure of LQC from LQG. More specifically, in LQC, which is a symmetry-reduced model of LQG, the different components of the Hamiltonian are treated as multiples of each other. This is not the case in the full theory of LQG. As such, mLQC-I and mLQC-II are two attempts to incorporate more aspects of LQG in LQC, by means of similar treatments of the Hamiltonian, with respect to the treatment given in LQG.

Furthermore, we applied the covariant order reduction method, by obtaining a reduced version of the full field equations of $f(R)$ gravity, in the sense that they are second-order equations and give solutions perturbatively close to GR, and deduced the modified Friedmann equations. This equation depends on the $f(R)$ function, which we chose to parametrize as $f(R) = R + \epsilon \varphi(R)$.
Motivated by the example given in Ref.~\cite{Sotiriou:2008ya}, in which a function $\varphi(R)$ was found, we applied this procedure to the mLQC-I and mLQC-II models.
The first model, mLQC-I, is divided in two branches, denoted by $b_{-}$ and $b_{+}$, providing the modified Friedmann equations given by Eqs.~\eqref{H2I} and~\eqref{H2I2}, respectively. The second model, mLQC-II, gives the modified Friedmann equation (\ref{H2II}). 
We then found a function $\varphi(R)$, such that Eqs.~\eqref{H2I}--\eqref{H2II} are the same as Eq.~\eqref{mFE2}, for $w=1$. As such, specific effective covariant actions were found, in the context of metric $f(R)$ gravity, which provide Eqs.~\eqref{H2I}--\eqref{H2II}.
Moreover, from Figs.~\eqref{fig1}--\eqref{fig4}, we are able to see that these effective actions satisfy $f'(R)>0$ and $f''(R)>0$. This is relevant since solutions with $f'(R)>0$ allow for a positive effective gravitational coupling and the condition $f''(R)>0$ avoids the Dolgov--Kawasaki instability~\cite{Sotiriou:2008rp}.
In principle, according to the order reduction method that we have used, all the solutions are valid only when condition (\ref{condition}) applies. With the exception of LQC, the calculations were managed by the software \textit{Wolfram Mathematica}.

A successful theory that combines GR and quantum mechanics is yet to be found. In order to predict and describe the beginning of the Universe, we need physical laws that are valid in that regime. If GR is the correct theory in describing the Universe in that period, then the singularity theorem would show that, in the beginning, the Universe was contained in a single point, with infinite density and infinite curvature. However, what the theorem really shows is that, in the beginning of time, the magnitude of the gravitational interaction was so strong, that quantum gravitational effects were, most likely, relevant. As such, it is expected that a quantum theory of gravity will allow for a proper description of the beginning of the Universe. In this regard, the $\Lambda$CDM model is incomplete and, for this reason, it is pertinent to attempt modifications of GR, in order to accommodate scenarios in which the Big Bang singularity is non-existent, such as the covariant effective actions determined in this work.
In a forthcoming paper we will extend our analysis to arbitrary values of $w$.

%%%%%%%%%%%%%%%%%%%%%%%%%%%%%%%%%%%%%%%%%%%%%%%%%%%%%%%%%%%%%%%%%
\section*{Acknowledgements}
%%%%%%%%%%%%%%%%%%%%%%%%%%%%%%%%%%%%%%%%%%%%%%%%%%%%%%%%%%%%%%%%%

DV acknowledges support from the {\it Istituto Nazionale di Fisica Nucleare} (INFN) ({\it iniziativa specifica} TEONGRAV). FSNL acknowledges support from the Funda\c{c}\~{a}o para a Ci\^{e}ncia e a Tecnologia (FCT) Scientific Employment Stimulus contract with reference CEECINST/00032/2018, and funding from the research grants No. UID/FIS/04434/2020, No. PTDC/FIS-OUT/29048/2017 and No. CERN/FIS-PAR/0037/2019. 

%%%%%%%%%%%%%%%%%%%%%%%%%%%%%%%%%%%%%%%%%%%%%%%%%%%%%%%%%%%%%%%%%

%%%%%%%%%%%%%%%%%%%%%%%%%%%%%%%%%%%%%%%%%%%%%%%%%%%%%%%%%%%%%%%%%

%%%%%%%%%%%%%%%%%%%%%%%%%%%%%%%%%%%%%%%%%%%%%%%%%%%%%%%%%%%%%%%%%
\end{document}